\documentclass[conference]{IEEEtran}

\usepackage{graphicx}
\usepackage{amssymb}
\usepackage{amsthm}
\usepackage[n,operators,advantage,sets,adversary,landau,probability,notions,logic,ff,mm,primitives,events,complexity,asymptotics,keys]{cryptocode}
\usepackage{listings}
\usepackage{float}
\usepackage[normalem]{ulem}
\usepackage{graphics} 
\usepackage{epsfig} 
\usepackage{bbm}
\usepackage{subfig}
\usepackage{color}
\usepackage{multirow}
\usepackage{enumitem}

\newtheorem{definition}{Definition}[section]

\usepackage{lineno}

\begin{document}

\title{Collaborative Homomorphic Computation on Data Encrypted under Multiple Keys}

\author{
\IEEEauthorblockN{Asma Aloufi\IEEEauthorrefmark{1}\IEEEauthorrefmark{4}, Peizhao Hu\IEEEauthorrefmark{1}}
\IEEEauthorblockA{\IEEEauthorrefmark{1}Rochester Institute of Technology, NY, USA}
\IEEEauthorblockA{\IEEEauthorrefmark{4}Taif University, Taif, Saudi Arabia \\ Corresponding authors: \{ama9000, Peizhao.Hu\}@rit.edu}
}

\maketitle

\begin{abstract}
Homomorphic encryption (HE) is a promising cryptographic technique for enabling secure collaborative machine learning in the cloud. However, support for homomorphic computation on ciphertexts under multiple keys is inefficient. Current solutions often require key setup before any computation or incur large ciphertext size (at best, grow linearly to the number of involved keys). In this paper, we proposed a new approach that leverages threshold and multi-key HE to support computations on ciphertexts under different keys. Our new approach removes the need of key setup between each client and the set of model owners. At the same time, this approach reduces the number of encrypted models to be offloaded to the cloud evaluator, and the ciphertext size with a dimension reduction from $(N+1)\times 2$ to $2\times 2$. We present the details of each step and discuss the complexity and security of our approach.
\end{abstract}

\section{Introduction}
\label{sec:intro}

Secure computation outsourcing allows the delegation of expensive computations to a resourceful cloud while preserving the privacy of user data~\cite{singh2016survey, shan2018practical}. Homomorphic encryption (HE) is a promising cryptographic technique that protects data in-transmission, at-rest, and in-use without decryption. These capabilities are important for enabling secure computation on private data in the cloud. 

Typically in well-known HE schemes, such as BGV~\cite{brakerski2012leveled}, B/FV~\cite{brakerski2012fully,Fan:2012aa}, GSW~\cite{gentry2013homomorphic}, homomorphic computation are performed on data that is encrypted under a single key pair. This leads to a weaker security model because all participants have to share the same key pair; that is, they can see each other's private data. Many cloud computing applications require stronger security, such that private data from different individuals is encrypted under different key pairs. This security model allows a group of users to contribute their encrypted data to a cloud evaluator for joint computations without giving away their privacy. We refer to such setting as \textit{secure collaborative computing}. 

\begin{figure}[h]
\centering\includegraphics[width=0.68\linewidth]{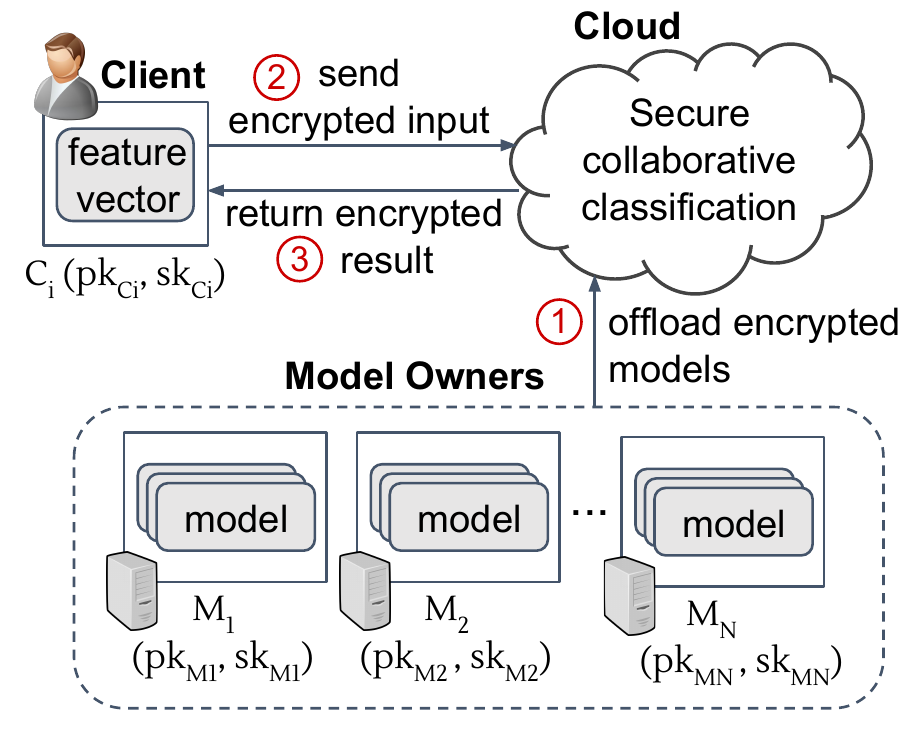}
\caption{Secure collaborative machine learning.}
\label{fig:secML}
\end{figure}

Collaborative machine learning~\cite{Hofmann2005aa} is an increasingly important application of secure collaborative computing because of the growing interesting in Machine Learning as a Service (MLaaS) in the cloud. A set of parties cooperate in training predictive models on their private datasets and perform secure classification for the given client inputs, as illustrated in Fig.~\ref{fig:secML}. For instance, in the application of medical diagnosis multiple hospitals and medical laboratories contribute their health data to train predictive models. In another example, credit agencies such as Equifax, Experian, and TransUnion, can jointly assess a customer's credit score but none of them are willing to give up their models. Joint datasets and models are more diverse and often contain features that help to achieve better accuracy. Operating on encrypted data helps to prevent incidents such as the Equifax data leak~\cite{Obrien2017}.
 
\subsection{Computing with multiple keys} 
Computing on ciphertexts that are encrypted under different key pairs can be tricky and inefficient. Threshold HE~\cite{asharov2012multiparty} can produce a joint key using system participants' keys based on the additive homomorphic property of secret keys. Before any computation, all participants cooperate to produce a joint key in a setup phase. Data will be encrypted using this joint key. At the end of the evaluation, the users must cooperate to decrypt the result using a multi-party computation (MPC) protocol. The security of this approach is based on the \emph{learning-with-error} assumption~\cite{Regev:2010aa} that hides the keys and a \emph{dishonest-majority} secret-sharing model~\cite{Lindell:2005aa}. In other words, no one individual can retrieve the result unless all are cooperating. The dishonest-majority model may not be efficient in the collaborative setting. Suppose we have $N$ model owners and $P$ clients, as illustrated in Fig.~\ref{fig:secML}, for every client we need to generate a joint key for this client and the group of $N$ model owners; that is, we will need to produce and maintain $P$ joint keys. This also means that each model owner has to provide $P$ copies of the encrypted model to the cloud.

As another solution, multi-key HE~\cite{lopez2012fly} supports homomorphic computation on ciphertexts encrypted under different keys without a joint key setup. Ciphertexts can be extended ``on-the-fly'' to a concatenation of participants' keys. The size, more specifically dimension, of an extended ciphertext increases with respect to the number of involved keys. The most efficient construction of MKHE~\cite{chen2017batched} is based on the BGV scheme~\cite{brakerski2012leveled}, where the ciphertext size increases linearly. For our collaborative machine learning scenario, an MKHE solution will require each ciphertext to be extended to $N+1$ different keys (i.e., the model owners keys plus the key of the requesting client) before any computation. The efficiency of the system is affected especially if the number of model owners is large because it proportionally increases the ciphertext size.

\subsection{Our approach}
In this paper, we propose a new approach that supports homomorphic computation over ciphertexts encrypted under multi-key and produces small ciphertexts. More specifically, because the group of $N$ model owners are likely to remain the same during the evaluation, we propose to generate a joint key using the threshold HE approach. Client's request can be dynamic; hence we adopt the idea of ``on-the-fly'' ciphertext extension to transform a ciphertext under the joint key to one under a concatenation of the model owner's joint key and the client's key. The same treatment is applied to the ciphertexts under the client's key. By purposefully combining the two approaches, we remove the requirement of generating a joint key for every client and the group of model owners. Also, we reduce the number of encrypted models from $P$ to $1$ and the dimension of the extended ciphertexts from $(N+1)\times 2$ to $2\times 2$. Note, these approaches are based on the BGV HE scheme; hence, the base ciphertext dimension is $1\times 2$. Table~\ref{tab:compareApproach} provides a summary of our comparison. In this analysis, we focus on the ciphertexts of the models because they are stored in the cloud and used in every evaluation. Also, we make another contribution in proposing a new decryption protocol based on the MPC protocol. 

\begin{table}[t]
\centering
\caption{A comparison of different approaches.}
\label{tab:compareApproach}
\begin{tabular}{|c|c|c|}
\hline
\textbf{Approach} & \multicolumn{1}{c|}{\begin{tabular}[c]{@{}c@{}}\# \textbf{of encrypted} \\ \textbf{model copies}\end{tabular}} & \textbf{Ciphertext size (dimension)}  \\ \hline
Threshold HE~\cite{asharov2012multiparty}   & $\textbf{P}$ & $1\times2$  \\ \hline
Multi-Key HE~\cite{chen2017batched} & $1$  & $(\textbf{N+1})\times2$ \\ \hline
This work & $1$  & $\textbf{2}\times2$ \\ \hline
\end{tabular}
\end{table}
 
\subsection{Organization}
The rest of the paper is organized as follows. An overview of the cryptographic techniques used in this paper is presented in Sec.~\ref{sec:prelim}. In Sec.~\ref{sec:approach}, we present the details and analysis of our approach. We discuss the  related work in Sec.~\ref{sec:related} and conclude the paper in Sec.~\ref{sec:conclusion}.

\section{Preliminaries}
\label{sec:prelim}

\subsection{Notations and definitions}

Given a vector $a = (a_1, \dots, a_n)$, we define $a[i] = a_i$ as the $i$-th element. Let $m\times n$ be the matrix dimensions and $B[i,j]$ be the $j$-th element of the $i$-th row. The dot product of the two vectors $a, b$ is denoted as $\langle a,b\rangle = \sum_{i=1}^{n}a[i]\cdot b[i]$, and the tensor product is denoted as $a\otimes b$.

For a security parameter $\lambda$, let $\Phi(x)=x^\eta+1$ be a cyclotomic polynomial where $\eta$ as a power of 2. Define the ring $R$ over the integers $R = \ZZ$, or over polynomials with integer coefficients $R=\ZZ[x]/(\Phi(x))$, which can be bounded by $q \in \ZZ$. Let $\chi$ be a Gaussian error distribution over $R_q=\ZZ_q$ and bounded by $\mathcal{B} \ll q$. The security of the HE schemes is based on the LWE assumption or its ring variant.

\begin{definition}[LWE~\cite{Regev:2010aa}]
Let $a$ and $s$ be uniformly sampled elements from $R_q = \ZZ_q$, and let $e\sample\chi$ be a sampled error term. The learning with errors (LWE) problem~\cite{Regev:2010aa} is to distinguish the pair of $(a_i, b_i=as + e)$ from any uniformly sampled pair $(a_i, b'_i) \sample R^2_q$. The LWE assumption is that LWE problem is computationally \textit{infeasible}. In another word, secret key $s$ is hidden by the element $a \in R_q$ with small noise.
\end{definition}

\begin{definition}[Ring-LWE~\cite{lyubashevsky2013ideal}]
Let $q$ and $t$ be two co-prime moduli where $q \gg t$, we define the ciphertext space $R_q=\mathbb{Z}_q[x]/(\Phi(x))$ and the plaintext space $R_t=\mathbb{Z}_t[x]/(\Phi(x))$. The RLWE assumption is simply the general LWE assumption but instantiated over the ring of polynomials $R_q$.
\end{definition}

\subsection{Key switching} 
\label{subsec:keySwitching}
This technique is applied after homomorphic operations to transform a ciphertext from one under the key $s$ to one under a different key $s'$. It is also called the \emph{relinearization} step~\cite{brakerski2012leveled} that reduces the dimension after each homomorphic multiplication and yield a normal ciphertext that is decryptable by the secret key $s'$. This transformation is accomplished with the aid of auxiliary information provided as evaluation key $\ek$ which encrypts $s$ under $s'$. To perform key switching, two essential functions are needed:

\begin{itemize}[leftmargin=*]
    \item[-]$\evalkgen(s,s')$: Given two keys $s\in R^k_q, s'\in R^2_q$, let $\beta=\floor{\log q}$ and compute the powers-of-2 of the old secret key $\Tilde{s} = \powersoftwo(s) = (2^0s, 2s,\dots, 2^\beta s)\in R_q^{k\beta}$. Sample $k\beta$ RLWE instances $(a_i, a_is'+te'_i)$ and output $\ek = \{(a_i, a_is'+te'_i+\Tilde{s}[i])\in R^2_q\}_{i=1,\dots,k\beta}$
    \item[-]$\keyswitch(\ek, c)$ Given a ciphertext $c\in R^k_q$ under $s$ and the evaluation key $\ek$, decompose the ciphertext to its binary such that $\Tilde{c} = \bitdecomp(c) = (u_0, \dots, u_{\floor{\log q}})$ where $c = \sum^{\floor{\log q}}_{i=0} (u_i 2^i)$ and output the new ciphertext as $c' = \sum_{i=0}^{k}(\Tilde{c}[i]\ek[i]) \in R^2_q$ which is encrypted under the new key $s'$. 
\end{itemize}

\subsection{Threshold HE}
\label{subsec:threshold}
Asharov~et~al.~\cite{asharov2012multiparty} proposed a threshold scheme extended from the BGV scheme~\cite{brakerski2012leveled}. Formally, given a set of $n$ public keys $\pk_i=(A, A s_i+t e_i)$ where $i\in \{1,n\}$, the element $A\in R_q$ is shared, $s_i$ is the underlaying private key. We generate a joint public key $\pk^*=(A, A\sum{s_i}+t\sum{e_i})$. Note that only the second component of the individual public keys is aggregated such that underlying partial secret keys $s_i$ are homomorphically added together under the RLWE assumption. If we add both components of the public keys, the joint key will be $\pk^*=(nA, As^*+te^*)$. Subsequently, a ciphertext encrypted under the joint key will be $c = (c_0, c_1) = (\gamma nA, \gamma As^*+te^*+ \mu)$. Decryption will fail because $(\gamma nAs^* \neq \gamma As^*)$. Two-round setup protocol is required to generate the joint evaluation key $\ek^*$, which is required for performing key switching after each homomorphic multiplication since its generation is trickier due to its fundamentally complex structure. To decrypt, each user contributes their partial key $s_i$ by computing $c_0s_i+te_i$. Then, all users collaboratively produce a component that contains the sum of all secret keys shares, that is $(c_0s^*+te^*)=(c_0\sum{s_i}+t\sum{e_i})$. Refer to the decryption process $\dec(c, \sk)=c_1-c_0 s^*$, the message can be decrypted correctly when we compute $c_1-(c_0s^*+te^*)$.

\subsection{The ring-GSW scheme}
\label{subsec:rgsw}
Chen~et~al.~\cite{chen2017batched} proposed a ring variant of the GSW scheme~\cite{gentry2013homomorphic} based on the RLWE problem. The ring-GSW (RGSW) operates in a ring of polynomials, and its plaintext is a ring element instead of a single bit. This scheme is used to generate the evaluation keys that are needed after each homomorphic operation in the multi-key BGV scheme as will be discussed in the next subsection. Here, we describe below the main functions of the RGSW scheme.

\begin{itemize}[leftmargin=*] 
\item[-]{$\mathsf{RGSW.}\setup(\lambda)$}: Given a security parameter $\lambda$, choose a modulus $t$ for plaintext space and a modulus $q$ for ciphertext space. Let $\beta = \floor{\log q} + 1$ be the bit length of the ciphertext modulus $q$. Define a $\mathcal{B}$-bounded error distribution $\chi$. Choose a cyclotomic polynomial $\Phi(x)=x^\eta+1$ that defines the polynomial ring $R=\ZZ[x]/(\Phi(x))$. Let $R_q = R / qR$ be a ring of polynomials of degree $\eta$ with integer coefficient in $\ZZ_q$. Output the public parameter $\param = (t,q,\beta,\chi,\mathcal{B},R_q)$.

\item[-]{$\mathsf{RGSW.}\kgen(\param)$}: Sample a small secret element $s'\sample \chi$ and an error vector $e \sample \chi^{2\beta}$. Choose a random vector $a \sample R^{2\beta}_q$ Set $b = s'a + te \in R^{2\beta}_q$. Output the secret key $\sk$ as $s = (1, -s') \in R^2_q$ and the public key $\pk$ as $(a, b) \in R^{2\beta\times2}_q$. Note that we need $\beta$ RLWE instances $(a,b)$ to encrypt $\beta$-bits of the secret key needed for evaluation key generation
\item[-]{$\mathsf{RGSW.}\enc(\pk, \mu)$}: Let $\mu \sample R_q$ be a message and $\pk = (a,b)$ be a given public key. Let $G$ be a \textit{gadget matrix} which is a diagonal matrix containing powers-of-2 $g= (1,2,4,\dots,2^{\beta-1})$ such that $G=(I_2, 2I_2,\dots, 2^{\beta-1}I_2)^T \in R^{2\beta\times2}_q$,  where $I_2$ is a $2\times 2$ identity matrix. To encrypt the message $\mu$, choose a random element $\gamma\sample \chi$ and two error terms $(e_0, e_1) \sample \chi^{2\beta\times2}$ and compute the ciphertext as: 
$$C = \gamma(a,b) + t(e_0, e_1) + \mu G \in R^{2\beta\times2}_q $$
Note that the propriety $sc = t(e_0,e_1) + \mu sG$ holds. 

\item[-]{$\mathsf{RGSW.}\encrand(\pk, \gamma)$}: Let $\gamma$ be the randomness used to encrypt a message $\mu$, and $\pk=(a,b)$ be a given public key. To encrypt the randomness, compute the powere-of-2 such that $\powersoftwo(\gamma)=(\gamma,2\gamma,\dots,2^{\beta-1}\gamma)$. Choose $\beta$ new random elements $\gamma' \sample \chi$ and two new error terms $(e'_0, e'_1) \sample \chi^{\beta}$. Compute the ciphertext as $F = (f_0,f_1)\in R^{\beta\times2}_q$, such that $f_0[i] = (b[i]\gamma'_i + te'_0[i] + \powersoftwo(\gamma)[i]) \in R_q$ and $f_1[i] = (a[i]\gamma'_i + te'_1[i]) \in R_q$. This encryption of randomness will be used to generate auxiliary information to help decrypting the extended ciphertexts.

\item[-]{$\mathsf{RGSW.}\extend(C_i, F_i, \{\pk_j; j=1,\dots,N\})$}: Given a set of $N$ public keys $(\pk_1, \dots, \pk_N)$ and a ciphertext $C_i$ encrypting a message $\mu$ under $\pk_i$, extend the ciphertext to multi-keys as follows
$$\Bar{C} =  \begin{bmatrix}
C_i & \dots & X_1 & \dots & 0 \\
\vdots & \ddots & \vdots &  & \vdots \\
0 & \dots & C_i & \dots & 0 \\
\vdots &  & \vdots & \ddots & \vdots \\
0 & \dots & X_N & \dots & C_i 
\end{bmatrix} $$

Where the component $X_j\in R^{2\beta\times2}_q$ is constructed using the public key $\pk_j$ and the encrypted randomness $F_j$ such that the $k$-th row is: $$X_j[k,:]= \bitdecomp(b_j[k])F_i \in R^{2}_q$$

The scheme supports multi-hop extension, which means that an extended ciphertext can be further extended to additional keys. With each extension, the gadget matrix $G$ is also extended to create a $2N\beta\times 2K$ matrix where the diagonal element is $G$. Note that the size of the extended ciphertext $\Bar{C}$ grows quadratically with the number of involved keys.
\end{itemize}

\subsection{The multi-key BGV Scheme} \label{subsec:mkbgv}
Chen~et~al.~\cite{chen2017batched} introduced the first multi-key BGV (MKBGV) scheme that is based on the RLWE assumption. The new scheme extends the BGV scheme~\cite{brakerski2012leveled} to a multi-hop MKHE setting (i.e., the extended ciphertext can be extended to further keys after the homomorphic evaluation). The scheme supports extending a ciphertext to a fixed number of keys. Similar to other RLWE-based schemes, the new multi-key BGV (MKBGV) scheme requires a generation of evaluation keys to perform key switching technique after each homomorphic operation. In the proposed scheme, the evaluation keys are generated as ring-GSW ciphertexts such that they can be extended to multiple keys which correspond to the extended ciphertext. We describe below the main functions of the proposed MKBGV scheme.  

\begin{itemize}[leftmargin=*]
\item[-]{$\mathsf{MKBGV.}\setup(\lambda, L)$}: Given the security parameter $\lambda$, a multiplicative depth $L$, a bound $K$ on the number of keys, define the ring of polynomials $R=\mathbb{Z}[x]/(\Phi(x))$ and the error distribution $\chi$ as described earlier. Choose the plaintext module $t$ that is co-prime with $L+1$ chosen decreasing ciphertext moduli $q_L\gg\dots\gg q_0$  for each level and set $\beta_l = \floor{\log q_l}+1$. For each level $l \in \{L,\dots,0\}$, uniformly sample an element $A_l\sample R_{q_l}^{2\beta_l}$. Finally, output $\param = (R, \chi, \{q_l, A_l\}_{l\in\{L,\dots,0\}}, t)$ as the public parameters for the scheme.

\item[-]{$\mathsf{MKBGV.}\kgen(\param)$}: To generate a key pair and the evaluation key for a user, perform the following steps.
\begin{itemize}
    \item[-] Randomly choose $L$ small elements $s'_l\sample\chi$ and define $s_l = (1, -s'_l)^T \in R^2_{q_l}$. Output the secret key as the set $\sk = \{s_l\}_{l\in\{L,\dots,0\}}$.
    
    \item[-] For each level $l\in\{L,\dots,0\}$, sample small noise $e_l\sample\chi^{2\beta_l}$ and define the pair $(b_l = A_ls'_l + te_l, A_l)\in R^{2\beta_l\time 2}_{q_l}$. Output the user's public key as the set of tuples $\pk = \{(b_l, A_l)\}_{l\in\{L,\dots,0\}}$.
      
    \item[-] For each level $l\in\{L,\dots,0\}$, generate the helper component $\ek'_l$ which is used later to generate the evaluation key for the performing homomorphic evaluation on extended ciphertexts. The component consists of two pairs of ring-variant of GSW ciphertexts which encrypt each bit of the user's secret key $s$ and encrypt the randomness used in the encryption. For each bit  $i=0,\dots,2\beta_l$, sample two random elements $r_{i,l}, r'_{i,l} \sample \chi$ and the error matrix $E_{i,l}=(e_{i,l,1}, e_{i,l,2})\in\chi^{2\beta_l\times 2}$. Compute the ring-GSW ciphertexts under the $(l-1)$-th public key element:
    \begin{equation*}
        \Theta_{i,l} = r_{i,l}(b_{l-1}, A_{l-1}) + tE_{i,l} + \mathtt{Powersof2}(s_l)[i]G
    \end{equation*}
    \begin{equation*}
        \Psi_{i,l} = r'_{i,l}(b_{l-1}, A_{l-1}) + tE'_{i,l} + \mathtt{BitDecomp}(s_l)[i]G
    \end{equation*}
    and the corresponding BGV encryptions of randomness $F_{i,l}$ (and $F'_{i,l}$), which encrypts each bit of the random element $r_{i,l}$ (and $r'_{i,l}$). 
    Finally, output the helper component for the evaluation key generated as $\ek'_l=\{(\Theta_{i,l},F_{i,l}),(\Psi_{i,l},F'_{i,l})\}_{i\in\{0,\dots,2\beta_l\}, l\in\{L,\dots,0\}}$
\end{itemize}

\item[-]{$\mathsf{MKBGV.}\enc(\pk, \mu)$}: Given a message $\mu\sample R_t$ and a user's public key $\pk$, perform the original BGV encryption function starting at level $L$. Later, as the ciphertext is evaluated, the level changes to a small $l\in\{L-1,\dots,0\}$

Sample error noises $e, e'\sample\chi$ and a random polynomial with binary coefficients $r\sample R_2$ and compute the ciphertext as $c = (c_0, c_1)\in R^{2}_{q_L}$ where $c_0 = rb_L+te+\mu$ and $c_1 = rA_L+te'$. Moreover, suppose that $S$ is an ordered set of all indexes of the users in which a ciphertext is encrypted under their secret keys. 

Output the fresh BGV ciphertext as the tuple $c = (c, S, L)$. We denote $[\cdot]$ as a regular BGV ciphertext encrypted under the public key $\pk$. Note that the output is fresh encryption; hence, it is starting at level $L$. However, this level changes as the ciphertext is evaluated. 

\item[-]{$\mathsf{MKBGV.}\extend(\{\pk_1,\dots,\pk_N\}, c)$}: To extend a BGV ciphertext at level $l$ to one encrypted under a set of $n$ users' keys, simply set the ciphertext $C$ as a concatenated $N$ sub-vectors $\Bar{c} = (c'_1|\dots|c'_N)\in R^{2n}_{q_l}$, such that $c'_i = c_i$ if the index $i\in S$, and $c'_i = 0$ otherwise. We denote $[[\cdot]]$ as an extended BGV ciphertext encrypted under the set of keys $\{\pk_1,\dots,\pk_N\}$.  

\item[-]{$\mathsf{MKBGV.}\evalkgen(\{\pk_1,\dots,\pk_N\}, \{\ek'_1,\dots,\ek'_N\})$}: This function generates an evaluation key $\ek_l$ for an extended ciphertext $\Bar{c} = (\Bar{C}, S, l)$ encrypted under $N$ public keys of the users in $S$. Given the set of public keys $\{\pk_1,\dots,\pk_N\}$ and the corresponding helper components $\{\ek'_1,\dots,\ek'_N\}$, generate the evaluation key as follows. First, for each user $i\in\{1,\dots,N\}$, extend each GSW encryption of the secret key $s_{i,l}$, that is encrypted under the next level's key $s_{i,l-1}$, to the other set of keys $\{s_{j\neq i, l-1}\}_{ j\in\{1,\dots,N\}}$. Then, using each encrypted bit of $\Bar{s}_{l}$, homomorphically compute $\Bar{s}'_{l} = \Bar{s}_{l} \otimes \Bar{s}_{l} \in R^{4N^2}_{q_l}$. The output of this operation are bit-wisely encrypted $\mathcal{K}_{p, \zeta}$ encrypted under the concatenated key corresponding to the next level $\Bar{s}_{l-1}$. Finally, output the extended evaluation key as $\Bar{\ek}_l = \{\mathcal{K}_{p,}\}$

\item[-]{$\mathsf{MKBGV.}\eval(\Bar{c}, \Bar{c}')$}: Given two extended ciphertexts $\Bar{c}, \Bar{c}' \in R_{q_l}^{2N}$ encrypted under the same concatenated keys $\Bar{s}_l$, perform homomorphic addition as element-wise addition $\Bar{c}_{\mathtt{add}} = \Bar{c} + \Bar{c}' (\mod q_l)$ or the homomorphic multiplication as the tensor product $\Bar{c}_{\mathtt{mult}} = \Bar{c} \otimes \Bar{c}'(\mod q_l)$. After the homomorphic evaluation, perform $\mathsf{KeySwitch}$ technique using the generated evaluation key $\Bar{\ek}_l$ to generate a ciphertext under the next level's key $\Bar{s}_{l-1}$, and the $\mathsf{ModulusSwitch}$ technique to reduce the resultant noise by switching to a smaller ciphertext modulus $q_{l-1}$. 

\item[-]{$\mathsf{MKBGV.}\dec(\{\sk_1,\dots,\sk_N\},\Bar{c})$}: Given an extended ciphertext $\Bar{c}$ encrypted at level $l$ under the set $S$ of users' keys, and the secret key $\Bar{s}_l = \{s_{1,l}|\dots|s_{N,l}\}$ that is the concatenation of all the users' secret keys in the set, decryption is as $\langle\Bar{c},\Bar{s}_l\rangle = \sum^{N}_{i=1}\langle c'_i, s_{i,l}\rangle = te + \mu = \mu \mod t$. The decryption is correct since each secret key $s_{i,l}$ decrypts the corresponding sub-vector ciphertext $c'_i$. 
\end{itemize}
\section{Proposed Approach}
\label{sec:approach}
   
\subsection{System setting}
We describe our approach with a working example of collaborative evaluation of random forest for classification tasks. In the collaborative setting, as shown in Fig.~\ref{fig:secML}, a semi-honest cloud ``evaluator'' classifies a client based on the encrypted decision trees from the $N$ model owners. More formally, each model owner $M_i; i\in\{1,\dots,N\}$ encrypts their decision tree $\mathcal{T}_i$ that consists of a set of decision nodes and leaf nodes under their respective keys. At each decision node, a boolean function is securely evaluated using an encrypted threshold $y$ and a client's encrypted input $x$. The output is an encrypted bit $b$, which is used to perform a conditional branching to traverse to the leaf node that contains a class label. At the end of the decision tree evaluation, the evaluator performs a secure counting protocol to calculate the frequency of each unique class label. Note, we focus on random forest evaluation for clarity, but the proposed approach is applicable to other ML techniques, such as deep learning because it requires matrix operations on data encrypted under different keys.

Boolean function within each decision node is the fundamental unit within a decision tree. For clarity, we focus our description on evaluating a Boolean function and a conditional branching program in one of the decision nodes. Similar evaluation procedures follow for the rest of the decision nodes. Suppose this decision node has two leaf nodes with two distinct class labels $A, B\in\{0,1\}$. Let a client $\mathcal{C}$ holds an input bit $x\in\{0,1\}$ encrypted under his key $\pk_\mathcal{C}$. Suppose each model owner $M_i$ holds an one node decision tree represented as the polynomial $\mathcal{T}_i =  b_i \cdot A + (1-b_i)\cdot B$ and a threshold $y_i\in\{0,1\}$ encrypted under a joint key $\pk_\mathcal{M}$. Assume this decision node contains a Boolean function $b_i = 1(x \neq y_i) = x + y_i~(\bmod~ 2)$, the output bit $b_i$ is then used to evaluate the branching program for $\mathcal{T}_i$ and output a class label $A$ or $B$. Note, the output of $\mathcal{T}_i$ is $A$ if $x$ and $y_i$ differ, otherwise it outputs $B$. Let the random forest $\mathcal{F} = \{\mathcal{T}_1,\dots,\mathcal{T}_N\}$ be the collection of random forest outputs. The final output of $\mathcal{F}$ depends on the frequency of each unique class label.

\subsection{Current solutions}
In the collaborative setting in Fig.~\ref{fig:secML}, one can use a threshold HE scheme to support computation with different keys. Mainly, participants (i.e., model owners $\mathcal{M}_i$  and client $\mathcal{C}_i$ ) generate among them a joint key from their individual keys to encrypt their inputs and compute under this joint key. Let $P$ be the number of clients registered in the system. Before the computation, the set of $N$ model owners must have a key setup with each of their clients $\mathcal{C}_i$ to generate a joint key $\pk_{\mathcal{M},\mathcal{C}_i}$. This means the model owners generate in advance $P$ different joint keys $\{\pk_{\mathcal{M},\mathcal{C_1}},\dots,\pk_{\mathcal{M},\mathcal{C_P}}\}$. This approach also means that each model owner must prepare $P$ encrypted copies of their models to the cloud, one for each distinct joint key.

As an alternative, participants can encrypt their inputs under their individual keys, but extend them to additional keys using the MKHE scheme at evaluation. This way, the model owners delegate one encryption copy of their models to the cloud. When a client $\mathcal{C}_i$ requests an evaluation, each SWHE ciphertext $c=(c_0, c_1)$ is extended to one in the form $\Bar{c} = \{c_{M_1}|\dots|c_{M_N}|c_{\mathcal{C}_i}\}\in R^{2(N+1)_q}$ under the set of $N+1$ keys $\{\pk_{M_1},\dots,\pk_{M_N}, \pk_{\mathcal{C}_i}\}$, resulting in linear expansion of the ciphertext size.

Our proposed approach leverages both the threshold and multi-key HE techniques. There is no need to set up a joint key for each client with the model owners, and ciphertexts are extended only under \textit{two} different keys instead of $N+1$ keys resulting in a reduction in the ciphertext size.

\subsection{Our new approach}
\begin{figure*}[t]
\centering
  \subfloat[Key Setup]
 {\includegraphics[width=.31\textwidth]{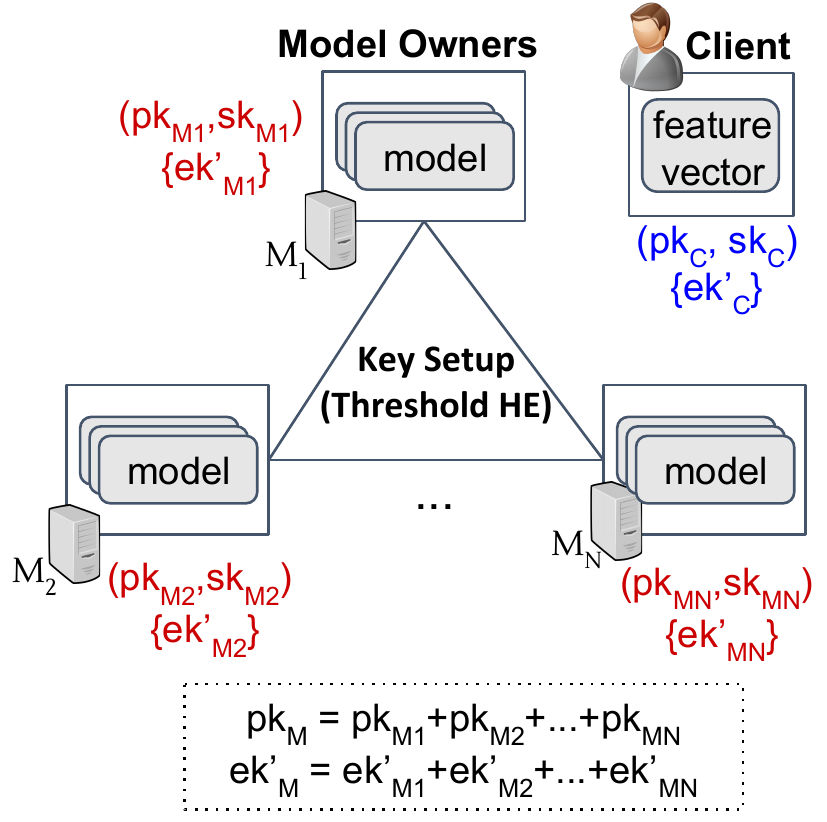}}\hfill
  \subfloat[Encryption and Evaluation]
 {\includegraphics[width=.34\textwidth]{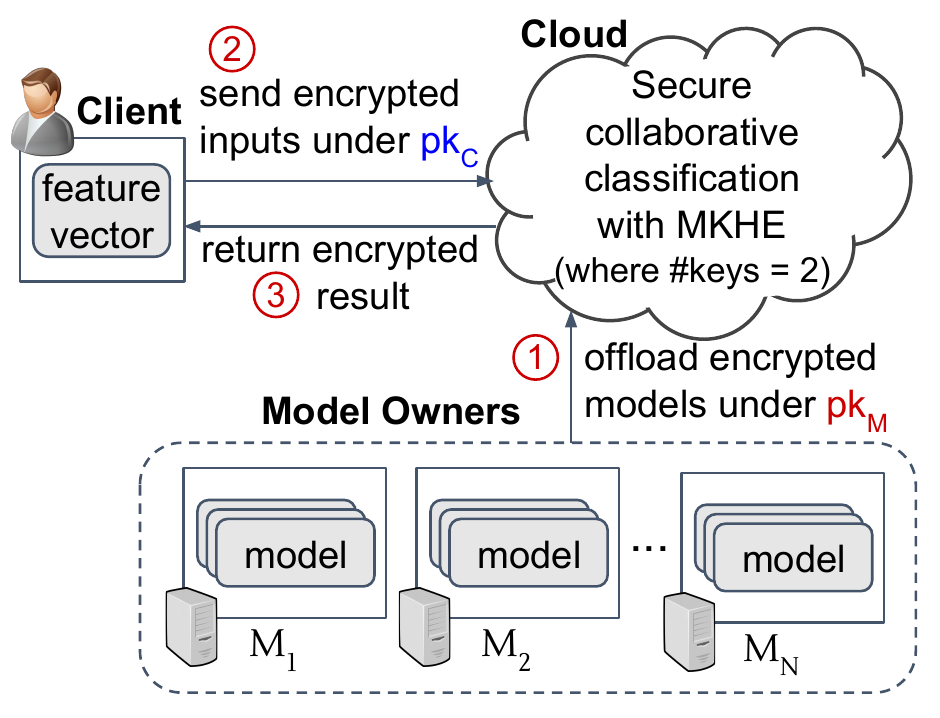}}\hfill
 \subfloat[Decryption Protocol]
 {\includegraphics[width=.34\textwidth]{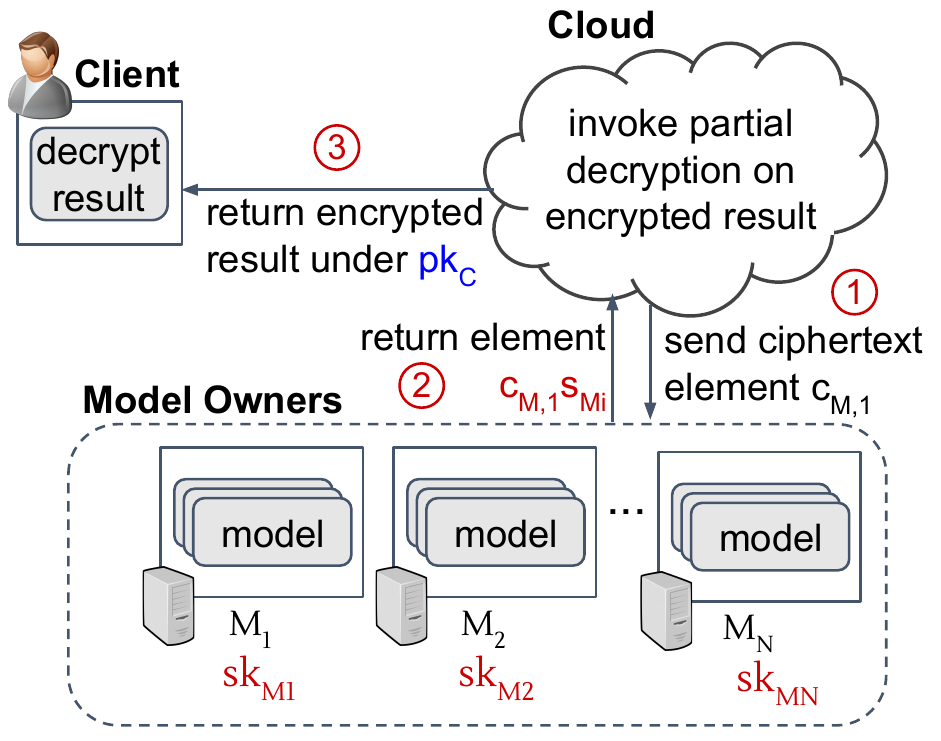}}
\caption{Illustration of the four main phases for secure collaborative evaluation protocol with multi-key support.}
\label{fig:four-phases}
\end{figure*}

Our new approach consists of four different phases: key setup, encryption, evaluation, and decryption. Figures~\ref{fig:four-phases} and~\ref{fig:phases} give an illustration and overview of these four phases.

\begin{figure*}
\hrule\hrule
\vspace{1.5em}
{\large\textbf{Setup phase.}}
\begin{LaTeXdescription} 
\small{
\item[Model owner $M_i \in \mathcal{M}$:] (1) Generate SWHE key pairs $(\pk_{M_i}, \sk_{M_i})$ and evaluation helper element $\ek'_{M_i}$ \\ 
(2) Generate with the other model owners the joint key $\pk_\mathcal{M} = \sum^{N}_{i=1}\pk_{M_i}$ and the evaluation helper element $\ek'_{\mathcal{M}} = \sum^{N}_{i=1}\ek'_{M_i}$.  \\
(3) Provide a threshold bit $y \in R_2$ and two possible output bits $A, B\in R_2$. \\
(4) Define a model as the polynomial $\mathcal{T}_i = b_i\cdot A + (1-b_i)\cdot B$, such that $b_i = 1(x\neq y_i)$.
\item[Client $\mathcal{C}$:] (1) Provide an input bit $x\in R_2$. \\
(2) Generate SWHE key pairs $(\pk_\mathcal{C}, \sk_\mathcal{C}))$ and the evaluation helper element $\ek'_\mathcal{C}$ \\ 
}
\end{LaTeXdescription}
{\large\textbf{Encryption phase.}}
\begin{LaTeXdescription}
\small{
\item[Model owner $M_i$:] (1) Encrypt the threshold and class labels bits $y_i, A, B$ using the joint public key $\pk_\mathcal{M}$, and output $[y_i]_\mathcal{M}, [A]_\mathcal{M}, [B]_\mathcal{M}$. \\
(3) Send $(\pk_{\mathcal{M}}, \ek'_{\mathcal{M}}, [y_i]_\mathcal{M}, [A]_\mathcal{M}, [B]_\mathcal{M})$ to the evaluator.  
\item[Client $\mathcal{C}$:] (1) Encrypt the bit $x$ using the SWHE public key $\pk_\mathcal{C}$, and output $[x]$. \\
(3) Send $(\pk_{\mathcal{C}}, \ek'_\mathcal{C}, [x])$ to the evaluator to start the evaluation. 
\item[Evaluator:] (1) Encrypt the value $1$ under the model owner's key $\pk_\mathcal{M}$, and output $[1]_\mathcal{M}$.
(2) Extend the ciphertexts $[y]_\mathcal{M}, [A]_\mathcal{M}, [B]_\mathcal{M}, [1]_\mathcal{M}$ from encryptions under $\pk_\mathcal{M}$ to ones under the extended SWHE key $\Bar{\pk} = \{\pk_\mathcal{M}, \pk_c\}$ using the extension function $\mathsf{MKBGV.}\extend$, and output $[[y]]=\{0, [y]_\mathcal{M}\}, [[A]]=\{0, [A]_\mathcal{M}\}, [[B]]=\{0, [B]_\mathcal{M}\}, [[1]]=\{0,[1]_\mathcal{M}\}$. \\
(2) Similarly, Extend ciphertexts $[x]_\mathcal{C}$ from encryption under $\pk_\mathcal{C}$, yielding $[[x]] = \{[x]_\mathcal{C}, 0\}$. \\
}
\end{LaTeXdescription}  
{\large\textbf{Evaluation phase.}}
\begin{LaTeXdescription}
\small{
\item[Evaluator:] (1) Evaluate $[[b]]=\mathsf{MKBGV.}\evaladd([[x]],[[y]])$; the result is $[[b]] = [[1]](x \neq y)$. \\
(2) Evaluate each decision tree $\mathcal{T}_i$ using $[[b]]$, and output $[[v_i]] = \mathcal{T}_i(x, y_i)$.\\
(3) Perform secure counting protocol among results, yielding $\mathcal{F}(x)=[[v_k]]$ where $z_k$ is maximum. \\ 
(4) Send the second ciphertext $[c]_\mathcal{M}$ of the encrypted result $\mathcal{F}(x)$ to each model owner $M_i$ to help with decryption.\\
}
\end{LaTeXdescription} 
{\large\textbf{Decryption phase.}}
\begin{LaTeXdescription}
\small{
\item[Model owner $M_i$:](1) Construct $\rho_i = c_1s_{M_i}$ from each ciphertext $[c]_\mathcal{M}=(c_0, c_1)$ using their secret share $s_{M_i}$. \\
(2) Add a large ``smudging'' error $te_i$ to the component, such that $\rho_i = c_1s_{M_i} + te_i$. \\ 
(3) Send this new component $\rho_i$ back to the evaluator.
\item[Evaluator:](1) Compute $\rho = \sum_{i=1}^N\rho_i = \sum_{i=1}^N(c_1s_{M_i} + te_{i})$. \\
(2) Send the encrypted result $\mathcal{F}(x)$ with the aggregated decryption component $\rho$ to the client.
\item[Client $\mathcal{C}$:](1) Decrypt the ciphertext $\mathcal{F}(X)$ that is encrypted under the extended key $\Bar{\pk}=\{\pk_\mathcal{C}, \pk_\mathcal{M}\}$ using the provided component $\rho$ and the client's own secret key $\sk_{\mathcal{C}}$ to obtain the final result.}
\end{LaTeXdescription} 
\vspace{0.5em}
\hrule\hrule
\vspace{0.5em}
\caption{An overview of secure decision tree evaluation phases using the proposed approach for multi-key support.}
\label{fig:phases}
\end{figure*}

\subsubsection{Key setup} 
In this phase, we generate key pair $(\pk_{M_i}, \sk_{M_i})$ for each model owner, and key pair $(\pk_{\mathcal{C}_i}, \sk_{\mathcal{C}_i})$ for each client. Moreover, we generate the corresponding evaluation helper elements $\ek'_{M_i}$ and $\ek'_{\mathcal{C}_i}$, which encrypts auxiliary information about the secret key. These elements are used to produce the evaluation key $\ek$, which is required to perform key switching after each homomorphic evaluation in the multi-key BGV scheme.

\paragraph{Model owners} 
In this system setting, the $N$ model owners do not change frequently; hence, we set up a joint key $\pk_\mathcal{M}$ once before the start of the protocol as shown in Fig.~\ref{fig:four-phases}(a). The joint key is generated, in a threshold manner, as the sum of the model owners' individual keys $\pk_\mathcal{M} = \sum_{i=1}^{N} = \pk_{M_i}$. It is used to encrypt the models before sending them to the cloud. Note that the encrypted model cannot be decrypted unless all the model owners collaborated since the corresponding secret key $\sk_\mathcal{M}$ is shared among them. The joint key can be revoked or updated, but this requires running the threshold key setup again and encrypting the models with the new joint key.

This threshold key setup differs from the one proposed by Asharov~\textit{et~al.}~\cite{asharov2012multiparty} because, at the end of their protocol, the evaluation key $\ek$ that is used in evaluation is directly generated. However, in our approach, we need first to generate the evaluation helper element $\ek'$ that can be extended later in the MKHE scheme described in Sec~\ref{subsec:mkbgv}. 

As mentioned, $\ek'$ must encrypt information on the bits of the secret key $s_\mathcal{M}$. The secret key $s_\mathcal{M}$ is shared among $N$ parties; therefore, we can generate the evaluation helper element by homomorphically adding the individual $\ek'_{M_i}$. The homomorphic addition can be achieved by using performing a fast full adder on the encrypted bits. The full adder algorithm contains a homomorphic multiplication, which is not the most efficient design. We briefly discuss optimization in Sec.~\ref{sec:analysis}. 

\paragraph{Clients} Each client $\mathcal{C}$ independently generates their own key pair $(\sk_\mathcal{C}, \pk_\mathcal{C})$. The output of the key generation step also includes a generated evaluation helper element $\ek'_\mathcal{C}$.

\subsubsection{Encryption} 
The model owner $M_i$ encrypts each threshold value $y_i$ and the class labels $A, B$ under the joint key $\pk_\mathcal{M}$. The model owner sends the encryptions and the evaluation helper element $(\pk_{\mathcal{M}}, \ek'_{\mathcal{M}}, [y_i]_\mathcal{M}, [A]_\mathcal{M}, [B]_\mathcal{M})$ to the evaluator. The evaluator then encrypts, under the joint public key $\pk_{\mathcal{M}}$, the value $1$ which is a part of the model $\mathcal{T}_i$.

On the other hand, the client encrypts the input bit $x\in\{0,1\}$ under their public key $\pk_\mathcal{C}$ and sends the encryption $[x]_\mathcal{C} \in R_q^2$ to the evaluator for evaluation.  

When the client $\mathcal{C}$ requests an evaluation, the evaluator first extends each ciphertext under the set of the two keys $\Bar{\pk} = \{\pk_\mathcal{M},\pk_\mathcal{C}\}$ as described in the extending algorithm $\extend(\{\pk_\mathcal{M}, \pk_\mathcal{C}\}, c)$ in Sec.~\ref{subsec:mkbgv}. The extended client's ciphertext is $[[x]] = \{[x]_\mathcal{C}, 0\}\in R_q^4$. Other ciphertexts are extended similarly to obtain $[[y]]=\{0, [y_i]_\mathcal{M}\}$, $[[A]]=\{0, [A]_\mathcal{M}\}$, and $[[B]]=\{0, [B]_\mathcal{M}\}$. 

\subsubsection{Evaluation}
For each model $\mathcal{T}_i$ in the joint model $\mathcal{F}=\{\mathcal{T}_1, \dots, \mathcal{T}_N\}$, compute the encrypted boolean value $[[b_i]]$ as the sum of the two extended ciphertexts $[[x]], [[y_i]]$, such that $[[b_i]] = [[x]] + [[y_i]] = \{[x]_\mathcal{C}, 0\} + \{0, [y_i]_\mathcal{M}\} = \{[x]_\mathcal{C}, [y_i]_\mathcal{M}\} \in R_q^4$. 

After that, the obtained boolean value is used to evaluate the model $\mathcal{T}_i(x) = [[v_i]] = [[b_i]] \cdot [[A]] + ([[1]] - [[b]]) \cdot [[B]]$, such that $[[v_i]] = \{[x]_\mathcal{C}, [y]_\mathcal{M}\} \cdot \{0, [A]_\mathcal{M}\} + (\{0, [1]_\mathcal{M}\} -$ $\{[x]_\mathcal{C}, [y]_\mathcal{M}\}) \cdot \{0, [B]_\mathcal{M}\}$.

The homomorphic multiplication of two extended ciphertexts $\Bar{c}, \Bar{c'} \in R^4_q$ is performed as a tensor product $\Bar{c}\otimes \Bar{c'} \in R^{16}_q$, which increases the dimension exponentially. The key switching technique is then applied with the help of the evaluation key $\ek$ to reduce the dimension, such that $\Bar{c}_{\mathtt{mult}} = \keyswitch(\ek, \Bar{c}\otimes \Bar{c'}) \in R_q^4$, and perform further homomorphic operations.

After evaluating each model $\{\mathcal{T}_1(x),\dots,\mathcal{T}_N(x)$, the results are aggregated using a secure count protocol (its implementation is omitted due to space limitations). The output of the protocol is the encryption of the class label with the highest count. 

\subsubsection{Decryption}
After the evaluation, the evaluator produces the result that is encrypted under the extended key $\Bar{\pk} = \{\pk_\mathcal{M}, \pk_\mathcal{C}\}$. To decrypt the extended ciphertext result $\Bar{c} = \{c_\mathcal{C}, c_\mathcal{M}\}$, we need the corresponding extended secret key $\Bar{s} = \{s_\mathcal{C}, s_\mathcal{M}\}$. Note that the secret key $s_\mathcal{M}$ is secretly shared among $N$ model owners, hence we propose the following decryption algorithm, $\dec(\Bar{c},\Bar{s}) = \langle c_\mathcal{C},s_\mathcal{C}\rangle + \sum^{N}_{i=1}\langle c_\mathcal{M}, s_{M_i}\rangle = (\mu'_\mathcal{C} + te_\mathcal{C}) + (\mu'_\mathcal{M} + te_\mathcal{M}) = \Bar{\mu} + t\Bar{e} \approx \mu~\bmod t$.

In the collaborative system design, the client will not be able to decrypt the result without model owners' secret shares of the key. Hence, the evaluator invokes a decryption protocol to obtain a decryption component needed to perform the part $\sum^{N}_{i=1}\langle c_\mathcal{M}, s_{M_i}\rangle$ from the decryption algorithm shown above. 

 In our protocol, the evaluator performs a partial decryption on the result such that it transforms it to a ciphertext under the client's key. As shown in Fig.~\ref{fig:four-phases}(c), the evaluator sends $c_{\mathcal{M}, 1}$ to all model owners. Upon receiving $c_{\mathcal{M}, 1}$, each model owner $M_i$ will construct $\rho_i = c_{\mathcal{M},1}s_{M_i}+te_{M_i}$, where $e_{M_i}$ is a large smudging noise, and return it back to the evaluator. After collecting the components from all model owners, the evaluator send the extended encrypted result $\Bar{c}$ to the client along with the aggregated value $\rho = \sum^{N}_{i=1}\rho_i = c_{\mathcal{M},1}s_{\mathcal{M}} + te_\mathcal{M}$. The client then computes: 
\begin{align*}
\mu' &= \langle c_\mathcal{C},s_\mathcal{C}\rangle + (c_{\mathcal{M},0} - \rho )
\\ &= (c_{\mathcal{C},0} - c_{\mathcal{C},1}s_{c}) + (c_{\mathcal{M},0} - c_{\mathcal{M},1}s_{\mathcal{M}} + te_\mathcal{M}) 
\\ &= \mu'_\mathcal{C} + te_\mathcal{C} + \mu'_\mathcal{M} + te_\mathcal{M}
\\ &= \Bar{\mu} + t\Bar{e} 
\\ &\approx \mu~\bmod t
\end{align*}

\section{Security and Complexity Analysis}
\label{sec:analysis}

\subsection{Security analysis}
Our proposed approach is secure in the semi-honest setting, where the system users follow the protocol specification and do not deviate from it. The semi-honest cloud evaluates on encrypted models and client inputs, which are protected under the semantic security of the underlying encryption scheme. 

The decryption protocol follows the dishonest-majority assumption where all the users are required to participate in decryption to retrieve the final classification result. Note that the cloud invokes a threshold decryption protocol to generate a decryption component from the model owners. This step generates the decryption component for the partial ciphertext $c_\mathcal{M}$ encrypted under $\pk_\mathcal{M}$, but the result remains encrypted under the client's key, who performs the final decryption.

\subsection{Complexity analysis}
We analyze the space and communication complexities of the proposed approach. The size of the extended ciphertext expands at most two concatenated ciphertext, i.e., a constant of size $2$. For communication, the one-time threshold key setup requires at most $N$ interactions between model owners. The collaborative evaluation of the models based on the client's input requires two interactions. The decryption protocol requires $2N$ interactions with the model owners.

\subsection{Optimizations}
The threshold key setup in our approach is performed in an MPC setting. While the design is secure, it is not efficient due to the full adder circuit that consists of homomorphic multiplication. Alternatively as an optimization, we can make a small security trade-off by using a trusted party. The trusted party generates a key pair $(\pk_\mathcal{M}, \sk_\mathcal{M})$, and the evaluation helper element $\ek'$ directly from the secret key $\sk_\mathcal{M} = s_{\mathcal{M}}$. To produce the secret shares of the secret key for each model owner $\{M_1, \dots, M_{N}\}$, we sample $N$ small secrets such that $\sk_{M_{i}} = (s_{\mathcal{M}} - \sum_{j=1}^{N}s_{M_j}); i\neq j$.

\section{Related Work}
\label{sec:related}

A common way to support multiple keys is to leverage the additive homomorphism of the key space. This property enables the establishment of a joint key from individual keys owned by individual users without the need for a trusted party. By direct aggregation of the users' public keys, we can effortlessly set up a $(n,n)$ threshold encryption scheme. Desmedt and Frankel~\cite{desmedt1998threshold} proposed a threshold version of ElGamal scheme~\cite{elgamal1985public} based on Shamir's secret sharing. Specifically, assume we have set of $n$ users where each with user $i$ has an independently generated key pair $(\pk_i = g^{s_i}, \sk_i = s_i)$. Then, the joint key is computed as $\pk^* = \prod_{i=1}^{n} \pk_i = \pk^{\sum^{n}_{i=1}s_i}$. A user can encrypt the data under the generated joint public key and compute using the homomorphic properties of the scheme. However, the decryption of the ciphertexts has to be performed by all $n$ users. Asharov~et~al.~\cite{asharov2012multiparty} proposed another threshold scheme extended from the BGV scheme~\cite{brakerski2012leveled}. Detail of this scheme was described in Section~\ref{sec:prelim}.

A more dynamic approach to support multiparty computation with multiple keys is multi-key HE. Lopez-Alt~et~al~\cite{lopez2012fly} introduced the first notion of MKHE schemes to support the homomorphic evaluation on ciphertexts encrypted under different keys. Its construction was based on NTRU. Many works~\cite{clear2015multi, mukherjee2016two, brakerski2016lattice, peikert2016multi, dodis2016spooky} followed after proposing constructions with different capabilities and security assumptions. Similar to threshold HE schemes, distributed decryption is required where the users collaborate to decrypt.

The first multi-key HE scheme that is based on the LWE problem was proposed by Clear and McGoldrick~\cite{clear2015multi}. It was then simplified by Mukherjee and Wichs~\cite{mukherjee2016two} who also built a general two-round MPC protocol on top of it. The basic scheme is constructed based on the GSW scheme~\cite{gentry2013homomorphic} and extended to support multiple keys. For an unbounded number of users, they can homomorphically compute on their individually encrypted inputs and output a ciphertext result encrypted under multiple keys. The scheme is described as \textit{single-hop}, which means that the ciphertext result cannot be extended to additional keys after being homomorphically evaluated. The decryption of the result can be retrieved by combining partial decryptions which were performed by each party. 

A more advanced MKHE enables \textit{multi-hop} key extension, which intuitively means that a homomorphically evaluated ciphertext can be further extended to additional keys. Peikert and Shiehian~\cite{peikert2016multi} proposed two (leveled) MKHE schemes basing their security on the standard LWE and its circular security assumptions. The core difference between the two schemes is the extension function which results in largely expanded ciphertexts in the first scheme, and small original GSW ciphertexts with large expanded public keys in the second scheme. A ciphertext can be expanded to one that is encrypted under a set of users' concatenated keys. The size of the ciphertext increases quadratically with the number of users.  They also proposed an alternative scheme, in which it yields small keys and large ciphertexts. More technically, the key pair is just GSW key, and helper information matrices are embedded in the ciphertexts instead of public keys. However, this design is less practical than their main proposed scheme since it is more likely that the number of homomorphic operations on the ciphertext exceeds the number of extended keys. 

Brakerski and Perlman~\cite{brakerski2016lattice} proposed a fully MKHE scheme based on the GSW scheme which realized extended ciphertexts that grows linearly with the size of included keys. The scheme applies on-the-fly bootstrapping technique to enable unlimited multiple evaluations on extended ciphertexts. Chen~et~al.~\cite{chen2017batched} proposed the first MKHE scheme that is based on RLWE. They extend the BGV scheme~\cite{brakerski2012leveled} to a multi-hop MKHE where the ciphertext size also grows linearly with the number of the bounded number of keys. The evaluation keys are generated as ring-GSW ciphertexts such that they can be extended to multiple keys which correspond to the expanded ciphertext.

Advancements have been proposed in the literature to extend homomorphic ciphertexts to multiple users' keys. The ciphertexts can be decrypted using the concatenated secret keys associated with public keys. Releasing secret keys to other parties presents security issues since a released secret key can be used to decrypt the user's other ciphertexts. The more secure approach used in proposed MKHE schemes is to hold a secure MPC decryption protocol where each of the involved users locally performs partial decryption of the ciphertext using his secret key. Yasuda~et~al.,~\cite{yasuda2018multi} proposed an alternative approach which applies Proxy Re-Encryption technique (PRE) to allow an extended ciphertext to be re-encrypted under the receiver's key such that it becomes decryptable by its secret key. Their work extends the Peikert-Shiehian MKHE scheme~\cite{peikert2016multi} to include two more functions, one for generating a re-encryption key, and the other is for performing the ciphertext re-encryption. 

\section{Conclusion}
\label{sec:conclusion}
We proposed a new approach that combines the threshold and multi-key HE to support collaborative computation on ciphertexts encrypted under different keys. In the collaborative machine learning setting, our proposed approach (i) removes the need of key setup between a client and the set of model owners, (ii) reduces the number of encryption of the same model, and (iii) reduces the ciphertext size with dimension reduction from $(N+1)\times 2$ to $2\times 2$. We presented the detail design of this approach and analyzed the complexity and security.

\bibliographystyle{unsrt}
\bibliography{ref.bib}

\begin{thebibliography}{10}

\bibitem{singh2016survey}
Saurabh Singh, Young-Sik Jeong, and Jong~Hyuk Park.
\newblock A survey on cloud computing security: Issues, threats, and solutions.
\newblock {\em Journal of Network and Computer Applications}, 75:200--222,
  2016.

\bibitem{shan2018practical}
Zihao Shan, Kui Ren, Marina Blanton, and Cong Wang.
\newblock Practical secure computation outsourcing: A survey.
\newblock {\em ACM Computing Surveys (CSUR)}, 51(2):31, 2018.

\bibitem{brakerski2012leveled}
Zvika Brakerski, Craig Gentry, and Vinod Vaikuntanathan.
\newblock {(Leveled)} fully homomorphic encryption without bootstrapping.
\newblock In {\em Innovations in Theoretical Computer Science Conference
  (ITCS)}, pages 309--325. ACM, 2012.

\bibitem{brakerski2012fully}
Zvika Brakerski.
\newblock Fully homomorphic encryption without modulus switching from classical
  {GapSVP}.
\newblock In {\em Advances in cryptology--crypto 2012}, pages 868--886.
  Springer, 2012.

\bibitem{Fan:2012aa}
Junfeng Fan and Frederik Vercauteren.
\newblock Somewhat practical fully homomorphic encryption.
\newblock https://eprint.iacr.org/2012/144/20120322:031216, March 2012.

\bibitem{gentry2013homomorphic}
Craig Gentry, Amit Sahai, and Brent Waters.
\newblock Homomorphic encryption from learning with errors:
  Conceptually-simpler, asymptotically-faster, attribute-based.
\newblock In {\em Advances in Cryptology--CRYPTO 2013}, pages 75--92. Springer,
  2013.

\bibitem{Hofmann2005aa}
Thomas Hofmann and Justin Basilico.
\newblock {\em Collaborative Machine Learning}, pages 173--182.
\newblock Springer Berlin Heidelberg, Berlin, Heidelberg, 2005.

\bibitem{Obrien2017}
Sara~Ashley O'Brien.
\newblock Equifax data breach: 143 million people could be affected.
\newblock
  \url{https://money.cnn.com/2017/09/07/technology/business/equifax-data-breach},
  Sep 2017.

\bibitem{asharov2012multiparty}
Gilad Asharov, Abhishek Jain, Adriana L{\'o}pez-Alt, Eran Tromer, Vinod
  Vaikuntanathan, and Daniel Wichs.
\newblock Multiparty computation with low communication, computation and
  interaction via threshold {FHE}.
\newblock In {\em Advances in Cryptology -- EUROCRYPT}, pages 483--501.
  Springer, 2012.

\bibitem{Regev:2010aa}
Oded Regev.
\newblock The learning with errors problem {(Invited Survey)}.
\newblock In {\em IEEE Conference on Computational Complexity}, pages 191--204,
  Cambridge, MA, June 2010.

\bibitem{Lindell:2005aa}
Yehida Lindell.
\newblock Secure multiparty computation for privacy preserving data mining.
\newblock In {\em Encyclopedia of Data Warehousing and Mining}, pages
  1005--1009. IGI Global, 2005.

\bibitem{lopez2012fly}
Adriana L{\'o}pez-Alt, Eran Tromer, and Vinod Vaikuntanathan.
\newblock On-the-fly multiparty computation on the cloud via multikey fully
  homomorphic encryption.
\newblock In {\em Proceedings of the forty-fourth annual ACM symposium on
  Theory of computing}, pages 1219--1234. ACM, 2012.

\bibitem{chen2017batched}
Long Chen, Zhenfeng Zhang, and Xueqing Wang.
\newblock Batched multi-hop multi-key fhe from ring-lwe with compact ciphertext
  extension.
\newblock In {\em Theory of Cryptography Conference}, pages 597--627. Springer,
  2017.

\bibitem{lyubashevsky2013ideal}
Vadim Lyubashevsky, Chris Peikert, and Oded Regev.
\newblock On ideal lattices and learning with errors over rings.
\newblock {\em Journal of the ACM (JACM)}, 60(6):43, 2013.

\bibitem{desmedt1998threshold}
Yvo Desmedt and Yair Frankel.
\newblock Threshold {C}ryptosystems.
\newblock In {\em Conference on the Theory and Application of Cryptology}, page
  307–315. Springer, 1989.

\bibitem{elgamal1985public}
Taher ElGamal.
\newblock A public key cryptosystem and a signature scheme based on discrete
  logarithms.
\newblock {\em IEEE transactions on information theory}, 31(4):469--472, 1985.

\bibitem{clear2015multi}
Michael Clear and Ciaran McGoldrick.
\newblock Multi-identity and multi-key leveled fhe from learning with errors.
\newblock In {\em Annual Cryptology Conference}, pages 630--656. Springer,
  2015.

\bibitem{mukherjee2016two}
Pratyay Mukherjee and Daniel Wichs.
\newblock Two round multiparty computation via multi-key fhe.
\newblock In {\em Annual International Conference on the Theory and
  Applications of Cryptographic Techniques}, pages 735--763. Springer, 2016.

\bibitem{brakerski2016lattice}
Zvika Brakerski and Renen Perlman.
\newblock Lattice-based fully dynamic multi-key fhe with short ciphertexts.
\newblock In {\em Annual Cryptology Conference}, pages 190--213. Springer,
  2016.

\bibitem{peikert2016multi}
Chris Peikert and Sina Shiehian.
\newblock Multi-key fhe from lwe, revisited.
\newblock In {\em Theory of Cryptography Conference}, pages 217--238. Springer,
  2016.

\bibitem{dodis2016spooky}
Yevgeniy Dodis, Shai Halevi, Ron~D Rothblum, and Daniel Wichs.
\newblock Spooky encryption and its applications.
\newblock In {\em Annual Cryptology Conference}, pages 93--122. Springer, 2016.

\bibitem{yasuda2018multi}
Satoshi Yasuda, Yoshihiro Koseki, Ryo Hiromasa, and Yutaka Kawai.
\newblock Multi-key homomorphic proxy re-encryption.
\newblock In {\em International Conference on Information Security}, pages
  328--346. Springer, 2018.

\end{thebibliography}

\end{document}